\documentclass[prd,twocolumn,amsmath,amssymb,floatfix,superscriptaddress]{revtex4}
 \usepackage{graphicx}
 \usepackage{epstopdf}
 \usepackage{amsmath,amssymb}
 \usepackage{bm}
 \usepackage{verbatim,color}
 \usepackage{multirow}

\DeclareFontFamily{OT1}{pzc}{}
\DeclareFontShape{OT1}{pzc}{m}{it}%
             {<-> s * [1.00] pzcmi7t}{}
\DeclareMathAlphabet{\mathscr}{OT1}{pzc}%
                                 {m}{it}

\newcommand{\be}{\begin{equation}}
\newcommand{\ee}{\end{equation}}
\newcommand{\bea}{\begin{eqnarray}}
\newcommand{\eea}{\end{eqnarray}}

\newcommand{\reffig}[1]{Fig.~\ref{fig:#1}}          

\newcommand{\reftab}[1]{Tab.~\ref{t:#1}}
       
\newcommand{\refsec}[1]{Sec.~\ref{sec:#1}}          
\def\ba#1\ea{\begin{align}#1\end{align}}

   \newcommand{\planck}{{\it Planck\/}}
      \newcommand{\wmap}{{\it WMAP\/}}
   \newcommand{\mnu}{M_{\nu}}

\begin{document}
 
   \title{Constraints on Neutrino Mass from Sunyaev--Zeldovich Cluster Surveys}
 
 \author{Daisy S. Y. Mak}
 \email{suetyinm@usc.edu}
\affiliation{Physics and Astronomy Department, University of Southern California, Los Angeles, California 90089-0484, USA}

\author{Elena Pierpaoli}
 \email{pierpaol@usc.edu}
\affiliation{Physics and Astronomy Department, University of Southern California, Los Angeles, California 90089-0484, USA}

\begin{abstract}
The presence of massive neutrinos has a characteristic impact on the growth of large scale structures such as galaxy clusters. We forecast on the capability of the number count and power spectrum measured from the ongoing and future Sunyaev-Zeldovich (SZ) cluster surveys, combined with cosmic microwave background (CMB) observation to constrain the total neutrino mass $\mnu$ in a flat $\Lambda$CDM cosmology. We adopt self-calibration for the mass-observable scaling relation, and evaluate constraints for the South Pole Telescope normal and with polarization (SPT, SPTPol), Planck, and Atacama Cosmology Telescope Polarization (ACTPol) surveys. We find that a sample of $\approx1000$ clusters obtained from the Planck cluster survey plus extra information from CMB lensing extraction could tighten the current upper bound on the sum of neutrino masses to $\sigma_{\mnu}=0.17$ eV at 68\% C.L. Our analysis shows that cluster number counts and power spectrum provide complementary constraints and as a result they help reducing the error bars on $\mnu$ by a factor of $4-8$ when both probes are combined. We also show that the main strength of cluster measurements in constraining $\mnu$ is when good control of cluster systematics is available. When applying a weak prior on the mass-observable relations, which can be at reach in the upcoming cluster surveys, we obtain $\sigma_{\mnu}=0.48$ eV using cluster only probes and, more interestingly, $\sigma_{\mnu}=0.08$ eV using cluster + CMB which corresponds to a $S/N\approx4$ detection for $\mnu\ge0.3$ eV. We analyze and discuss the degeneracies of $\mnu$ with other parameters and investigate the sensitivity of neutrino mass constraints with various surveys specifications.

\end{abstract}

\maketitle

\section{Introduction}
Measuring masses of neutrinos is one the major goals of particle physics and cosmology. While atmospheric and solar neutrino oscillation experiments are sensitive to neutrino flavor, mixing angle, and the mass difference among different species, cosmological data are instead more sensitive to the absolute mass scale $\mnu=\sum m_\nu$. In fact, the most stringent upper bound of the total neutrino mass is coming from CMB and large scale structures since massive neutrinos leave detectable imprints throughout the history of the universe. Most recently,~\cite{Planck2013XVI} obtained $\mnu<0.23$ eV at $95\%$ C.L. by combining CMB data and BAO from Sloan Digital Sky Survey (SDSS)

In this work, we explore the prospects of employing ongoing and future galaxy cluster surveys detected by the SZ effect in constraining neutrino masses. Galaxy clusters are in principle a powerful tool for probing neutrino properties. 
Neutrino becomes nonrelativistic after the epoch of decoupling if its mass scale is smaller than $\mathcal{O}(0.1)$ eV. 
The relativistic behavior of neutrinos, as opposed to cold dark matter, causes the suppression of   matter perturbations  on small scales with respect to the case in which neutrinos are massless  and all dark matter is cold.
The presence of massive neutrinos  affects the growth rate of perturbations in the linear regime, and, as a consequence, the shape of the matter power spectrum and cluster abundance.
Current measurements from X-ray cluster surveys obtained a tight upper limit of $\mnu<0.33$ eV~\cite{Vikhlinin09} by combining measurements of Chandra X-ray observations of galaxy clusters, CMB from WMAP 5 year data, BAO and type 1A supernova (both from HST Key Project). Similarly, measurement from galaxy power spectrum from the SDSS-III BAO survey + CMB + SN found $\mnu<0.34$ eV~\cite{Zhao2013}.

Subsequently, several works were dedicated to discuss the prospects of utilizing large future surveys of large scale structures (galaxy or galaxy clusters) in different wavelengths (e.g.~\cite{Wang2005,Pritchard2008,Joudaki2012,Carbone2011,Carbone2012,Shimon2011,Shimon2012,Burenin2013}). These works showed that constraints of neutrino mass depend on assumptions of the underlying cosmology (e.g. inclusion of dark energy or flatness), cluster physics, and the use of external priors (e.g. CMB lensing extraction). Here we revisit the analysis to forecast the constraint of the total neutrino mass, in the framework of flat $\Lambda$CDM universe and, like past constraints, the standard scenario with only three neutrino species. We use cluster abundance and power spectrum as the observables that will be obtained from various SZ cluster surveys: the Planck, ACTPol, SPT, and SPTPol cluster surveys. These surveys are very promising and, in the next couple of years, will provide large samples of mass selected clusters out to high redshift. With respect to previous works~\cite[e.g][]{Shimon2011,Shimon2012} which also employ SZ cluster surveys, we provide a more realistic survey specifications to characterize the cluster detection and include the self-calibration to characterize the uncertainties of the mass-observable relations. We also discuss the degeneracy of the neutrino mass with dark energy, which is lacking in previous studies, and compare the strength of cluster probes with CMB on the constraining power of neutrino mass.

The paper is organized as follows. In~\refsec{theory} we discuss the effects of $\mnu$ on the large scale structures. In~\refsec{analysis} we present the methodology which includes the description of  the future SZ cluster samples and the Fisher matrix formalism. The main results are presented~\refsec{result} and discussed in~\refsec{discussion}. Finally, a conclusion is presented in~\refsec{conclusion}.

\section{Impact of neutrino masses on growth of the large scale structures}
\label{sec:theory}

The presence of massive neutrinos mildly affects expansion history but significantly impacts the  growth of structure through free-streaming. 
Fluctuations on comoving scales that enter the horizon when neutrinos are still relativistic may be reduced in amplitude because neutrinos would tend to leave the perturbation. This effect, that is neutrino mass dependent,  typically occurs on length scales below the 
free--streaming scale: $l_{\rm fs}=1/k_{\rm fs}=1/(1.5\sqrt{\Omega_M h^2/(1+z)}) ({\rm eV}/\mnu) {\rm Mpc}$. Thus, the growth of any structure that have scale smaller than $l_{\rm fs}$ will be less efficient. A smaller neutrino mass increases the free-streaming scale, but also reduces the neutrino fraction with respect the total amount of dark matter, mitigating the overall suppression.

As a result of these dependences measurements of the large scale structures such as cluster number counts and power spectrum can be used to place constraints on neutrino masses.

The late-time evolution of perturbations in a $\Lambda$CDM cosmology with massive neutrinos can be accurately described by the product of a scale dependent growth function and a time dependent transfer function. For example,~\cite{Lesgourgues2006b} derived a reasonable approximation to the analytical expression of the transfer function for small scales.
In this work, we employ the transfer function determined numerically from CAMB~\cite{Lewis2000} which provides precise estimate on the matter power spectrum and include non-linear effects at large--$k$ limit in which the analytical expressions fail to give an accurate estimate.

\section{Analysis}
\label{sec:analysis}
Our analysis closely follows the treatment of~\cite{Mak2012b}, here we only outline the method and refer the readers to~\cite{Mak2012b} for details. For cluster abundance and clustering, we use the results of numerical simulations from~\cite{Tinker2008} for the cluster mass function $n(M,z)$, and~\cite{Tinker2010} for the halo bias.

\subsection{Cluster survey}
We consider four upcoming SZ cluster surveys studied in~\cite{Mak2012b}: the \planck\ survey, the South Pole Telescope normal and polarization survey (SPT and SPTPol respectively), the Atacama Cosmology Telescope polarization survey (ACTPol).
Each of these surveys has different specifications for the selection threshold, i.e. $M_{\rm lim}(z)$, and their properties are summarized in~\reftab{survey}.

Briefly, for the \planck\ survey we adopt a flux limit of $Y_{200,\rho_c}\ge2\times10^{-3}{\rm arcmin^2}$~\cite{Melin2006}, where $Y_{200,\rho_c}$ is the integrated comptonization parameter within the radius enclosing a mean density of 200 times the critical density. This corresponds to a $5\sigma$ detection threshold and would yield $\sim1000$ clusters. 

For the SPT survey, i.e. single frequency at 150 GHz, we employ the calibrated selection function of the survey by~\cite{Vanderlinde2010} and adopt a detection threshold at $5\sigma$. This yields $\sim500$ clusters . The SPTPol has an increased sensitivity at 150 GHz than the normal survey and we account for this, following previous work, by scaling the mass limits by a factor of 3.01/5.95. The expected number of clusters is $\sim1000$.

For the ACTPol survey, we include clusters with $M_{\rm 200, \hat{\rho_c}}>5\times10^{14} M_{\odot} h^{-1}$ (Sehgal 2011, private communication) which corresponds to a $90\%$ completeness. This straight mass cut result in $\sim500$ clusters.


We construct cluster sample for the \planck\ survey in the redshift range $0<z<1$. We impose a lower cut $z_{\rm cut}=0.15$ for the SPT, SPTPol, and ACTPol survey. Currently, the SPT team is setting a low redshift cut at $z_{\rm cut}=0.3$ in their released cluster sample, due to difficulties in reliably distinguishing  low-redshift clusters from CMB fluctuations in single frequency observations. Nevertheless, with upcoming multi-frequency observations, a lower cut $z_{\rm cut}=0.15$ will likely be attained. We therefore apply this cut in our work. 


\begin{table}
\caption{Properties of SZ cluster survey}
\begin{center}
\begin{tabular}{ccc}
\hline\hline
   Survey & Area (sq. deg)& No. of clusters \\
\hline
\planck\ & 30000   & 1000 \\
SPT & 2500 & 500 \\      
SPTPol & 625 & 1000 \\
ACTPol & 4000 & 500 \\
\hline
\end{tabular} 
\end{center}
\label{t:survey}
\end{table}

\subsection{Fisher matrix forecasts and cosmological parameters}
We estimate the constraints on cosmological parameters by applying the Fisher matrix formalism to future SZ cluster surveys. This approach can best approximate the likelihood when the fiducial model is close to the true, underlying model and the likelihood is close to gaussian. Typically, the gaussian approximation is more accurate, and the use of the Fisher matrix better justified, when the likelihood is peaked and the parameter in hand has little degeneracies with other parameters.  In order to achieve this goal, the use of external priors can be beneficial.

For example,~\cite{Perotto2006} noted that the CMB power spectra  likelihood function for the neutrino mass differs from the gaussian case  due to strong parameter degeneracies, particularly for models with many parameters. These authors suggested the use of  CMB lensing extraction information in order to  sharpen the likelihood and make it  better approximated by  Gaussian. We will adopt the same strategy, as described below.

The Fisher matrix for the cluster number counts and power spectrum is described in detail in~\cite{Mak2012b}. Similarly, for our main results, we here consider self-calibration to account for the uncertainties of the observed cluster mass. We add the \planck\ CMB  lensing extraction (LE) that is considered to be a very promising way to constrain neutrino mass (e.g.~\cite{Lesgourgues2006b,Namikawa2010}). The CMB anisotropies obey Gaussian statistics in the absence of weak lensing, and therefore they are fully described by the temperature and polarization power spectrum. Weak lensing, however, introduces non-gaussianity in both the temperature and polarization anisotropies~\cite{Seljak1996,Bernardeau1997}. Therefore, extracting the lensing information from CMB (e.g. using quadratic estimators~\cite{Hu2001,Hu2002c,Hirata2003,Okamoto2003}) would provide the lensing potential and delensed CMB anisotropies, and hence extra information to the Fisher matrix. In the following, we refer to the Fisher matrix results obtained from CMB lensing extraction as the CMB LE. As shown in~\cite{Lesgourgues2006a}, CMB LE is useful in providing strong neutrino mass constraints and potentially breaking of the major neutrino mass degeneracies with other parameters~\cite{Perotto2006}. While very promising, the exploitation of higher order statistics may suffer from subtle ways from the effect of galactic and extragalactic contaminants. For this reason, we also consider constraints coming from the CMB power spectrum only (with lensing) when combining probes with cluster' s ones.

We note that the latest \planck\ results were released during the preparation of this work. They derive a tight upper limit of $\mnu\le0.93$ eV when using CMB data alone and $\mnu\le0.23$ eV when further combined with BAO data. Nevertheless, these limits use information from polarization of the \wmap\ data and not from the \planck\ data itself (the \planck\ CMB polarization data will be employed in the next data release). Therefore instead of using these numbers as priors on $\mnu$ constraints, we derive our \planck\ CMB prior  that takes into account the \planck\ polarization information which is believed to be better than that from \wmap. Thus this prior should be considered as the self-contained and improved one than the current constraint in~\cite{Planck2013XVI}.

We adopt a spatially flat $\Lambda$CDM model as the fiducial model. The set of parameters included in our analysis is $(\Omega_b h^2, \Omega_M h^2, \Omega_\Lambda, \mnu, n_s, \sigma_8, w_0, w_a)$. The fiducial values are adopted from the best fit flat $\Lambda$CDM model from WMAP 7yr data, BAO and $H_0$ measurements~\cite{Komatsu2011}: $\Omega_b h^2=0.0245$, $\Omega_M h^2=0.143$, $\Omega_\Lambda=1-\Omega_M=0.73$, $\mnu=0.3$ eV, $n_s=0.963$, $\sigma_8=0.809$, $w_0=-1$, $w_a=0$. 

As proposed in~\cite{Majumdar2003, Majumdar2004}, we can use cluster surveys to constrain the mass observable relation by considering self-calibration, hence taking into account the systematic errors of the SZ surveys due to uncertainties in observed cluster mass. In this work, we follow~\cite{Lima2005} to introduce four nuisance parameters, $B_{M0}$, $\alpha$, $\sigma_{\ln M,0}$, $\beta$, that specify the magnitude and redshift dependence of the fractional mass bias $B_{M}(z)=B_{M,0}(1+z)^\alpha$ and the intrinsic scatter $\sigma_{\ln M}(z)=\sigma_{\ln M,0}(1+z)^\beta$. We adopt fiducial values of $B_{M0}=0$, $\alpha=0$, $\sigma_{\ln M,0}=0.1$, $\beta=0$, hence corresponding to zero mass bias and $10\%$ intrinsic scatter. In deriving the main results, we will not make any assumption on the four nuisance parameters and leave them free to vary. We discuss the impact of this assumption in~\refsec{nusiance}.

\section{Results}
\label{sec:result}
\subsection{Cluster number count and power spectrum }
\reftab{results} summarizes the neutrino mass constraints from the Fisher matrix analysis for \planck\ CMB (with and without LE), cluster number counts, and power spectrum for the four cluster surveys. Constraints of $\mnu$ from cluster number counts alone are better than power spectrum ones, however, each of them is very weak when considered separately, with $\sigma_{\mnu}>4$ eV. When combining information from both probes, the constraints are improved significantly by a factor of $4-8$. The best case is obtained from the \planck\ cluster survey with $\sigma_{\mnu}=0.94$ eV, whereas the constraints from other surveys are a factor of two worse.  


\subsection{Cluster probes + CMB}
Adding the \planck\ CMB priors breaks degeneracies (see~\refsec{deg}) and improves the constraints (number count or power spectrum alone) further by a factor of $>4$ (without LE) and $>5$ (with LE).  When including all the information but LE, i.e. count + power spectrum + CMB, we find the best constraint comes from the \planck\ and ACTPol cluster survey with $\sigma_{\mnu}=0.23$ eV. This is $80\%$ better than that obtained from \planck\ CMB alone ($\sigma_{\mnu}=0.41$ eV). Including CMB priors also shrinks the difference in $\sigma_{\mnu}$ among different surveys in which it is now $\sigma_{\mnu}=0.23-0.30$ eV.  Similar results are obtained when we add the CMB LE and the best constraint is $\sigma_{\mnu}=0.17$ eV. This suggests that the improvements in $\sigma_{\mnu}$ are mainly driven by CMB information. We note that a perfect cleaning of all the astrophysical foregrounds is assumed when computing the CMB Fisher matrix in this work. Foreground contamination dominates at small angular scales (e.g. $l\ge1000$) and would introduce extra non-gaussianity and spoil the lensing extraction process~\cite{Cooray2005}. Nevertheless,~\cite{Lesgourgues2006a} found that the effect of no foreground subtraction in \planck\ CMB (with and without LE) only degrades the $\mnu$ constraint marginally (by $9\%$). Therefore, our results that involve CMB information can be considered to be robust against foreground contamination. 

We repeat the analysis with a fiducial $\mnu=0.1$ eV instead to investigate the effect on the constraint with less massive neutrinos. The results are very close (within $15\%$) to those for $\mnu=0.3$ eV when using cluster probes only, and are almost the identical when CMB priors are added.


\begin{table*}
\caption{Marginalized $1\sigma$ errors on $\mnu$ (in units of eV).}
\begin{center}
\begin{tabular}{cc|l | ccc  | ccc | ccc}
\hline\hline
    & & & \multicolumn{3}{|c|}{no prior } & \multicolumn{3}{|c|}{ + CMB prior} & \multicolumn{3}{|c}{ + CMB LE prior} \\
    CMB & CMB LE & Survey & NC & $P(k)$ &Comb & NC & $P(k)$ &Comb  & NC & $P(k)$ &Comb\\
\hline
\multispan4\hfil fiducial $\mnu=0.3$ eV\hfil\cr
\hline
\multirow{5}{*}{0.41} & \multirow{5}{*}{0.21} & Planck & 4.06&  7.83 &  0.94 &  0.29 &   0.29 & 0.23 & 0.20 &              0.20 &           0.17\\
         & &ACTpol  &           6.04       &             11.73 &             3.33 & 0.29 &              0.29 &           0.23 &           0.20 &              0.20 &           0.17\\              
          & &   SPT  &            12.44       &             12.45 &             2.12 &            0.33      &             0.31 &            0.30 &            0.21      &             0.20 &            0.20 \\
           &&  SPTpol  &            12.59       &             7.81 &             1.79 &            0.32       &             0.30 &             0.28&            0.21       &             0.20 &             0.19 \\
\hline
\multispan4\hfil fiducial $\mnu=0.1$ eV\hfil\cr
\hline
\multirow{5}{*}{0.52} & \multirow{5}{*}{0.19} &Planck  &            5.09       &             19.98 &             0.77&           0.37 &              0.42 &           0.23 &           0.18 &              0.19 &           0.15\\
         &&    ACTpol  &           17.97       &             48.63 &             2.83 &           0.43       &             0.38 &             0.26   &           0.19       &             0.18 &             0.16 \\         
         &&   SPT  &            6.56       &             31.39 &             1.78 &            0.43      &             0.39 &            0.28&            0.19      &             0.18 &            0.17 \\
         &&    SPTpol  &            12.59       &             7.81 &             1.79 &            0.32       &             0.30 &             0.28 &            0.21       &             0.20 &             0.19 \\

\hline
\end{tabular} 
\end{center}
\label{t:results}
\end{table*}

\subsection{Self-calibration and uncertainty of nuisance parameters}
\label{sec:nusiance}
The dominant systematic errors for SZ derived constraints are the uncertainties in the mass observable relation due to structure and evolution of clusters. We can ask how much could be gained by eliminating such uncertainties. For example, we can expect some external constraints on the nuisance parameters by using detailed studies of individual clusters or combining different information from optical, weak lensing, X-ray and SZ measurements. To estimate the effect of self-calibration of systematic uncertainties on the neutrino mass constraints, we repeat the forecasts with different priors on the four nuisance parameters as summarized in~\reftab{result_prior}.

We first discuss the results when applying a "weak" prior, i.e. using current knowledge on the calibration on the mass proxies $\Delta \sigma_{M,0}=0.1$, $\Delta \beta=1$, $\Delta B_{\rm M,0}=0.05$, $\Delta \alpha=1$. In the case of cluster count + power spectrum, the $1\sigma$ error reduces marginally for SPT and SPTPol, but significantly (by a factor of two) for \planck\ and ACTPol. This results in $\sigma_{\mnu}=0.48$ eV for the \planck\ cluster survey which is competitive with the CMB only constraint. In the case of adding the CMB (with and without LE) priors, the $1\sigma$ errors generally reduce by a factor of two and resulted in, for the best case as obtained by the \planck\ survey, $\sigma_{\mnu}=0.08$ eV, which corresponds to a $S/N\approx4$ detection for $\mnu\ge0.3$ eV

Similar results are obtained when applying a "strong prior", i.e. the four nuisance parameters are held fixed at their fiducial values, which is equivalent to assuming a perfect knowledge of cluster true masses. The constraints are improved significantly by $66-236\%$ in the case of cluster count + power spectrum, and a factor $2-3$ when the CMB priors are further added. The best constraint is, again with the \planck\ cluster survey, $\sigma_{\mnu}=0.07$ eV which is a relative marginal improvement with respect to the weak prior case. 
While it is unrealistic to have perfect knowledge on the mass observable relations, one can achieve similar scenario by restricting the analysis to a relatively small subset of clusters for which follow up observations are available. This would ensure a sample with well calibrated mass proxies. For example, it has been shown in~\cite{Vikhlinin09} that the ability to constrain dark energy parameters from a small sample of $\approx50$ well calibrated X-ray clusters is comparable to a larger sample of $\approx10000$ optical clusters (e.g. SDSS~\cite{Rozo2007}). 

Unlike other parameter constraints (e.g. non-Gaussianity with galaxy clusters~\cite{Mak2012b}), the results of the weak prior are sufficiently close to the those from the strong prior. The prospect of achieving the weak prior conditions is promising, e.g. clusters detected in weak lensing measurements or a subsample of objects having extensive multi-wavelength follow-up. 
Therefore the cluster probes are good enough to provide interesting $\mnu$ constraint even without perfect knowledge of the scaling relations.

As a final remark, we would like to compare our count + CMB result with~\cite{Shimon2012} which similarly presented $\mnu$ constraints assuming perfect knowledge of cluster mass and used \planck\ cluster count + CMB. Our result ($\sigma_{\mnu}=0.17$ eV) is a factor of 2.8 worse than that obtained in~\cite{Shimon2012}. We note that the discrepancy is due to the different assumption on the total number counts: $\approx6000$ in~\cite{Shimon2012} and $\approx1000$ in this work for the \planck\ survey if a $5\sigma$ survey detection limit is assumed. Our estimate is based on the conservative assumption that ensures high level of completeness ($90\%$) and realistic mass limits that vary at different redshifts, while~\cite{Shimon2012} assumed a constant and lower mass threshold. 

\begin{table*}
\caption{Fractional improvement $\frac{\sigma_{\mnu, no}}{\sigma_{\mnu, weak/strong}}$ with various priors (see~\refsec{nusiance}). The values of $\sigma_{\mnu,no}$ of the corresponding cases are those from~\reftab{results}. A fiducial $\mnu=0.3$ eV is assumed. The best case is obtained by the \planck\ survey, $\sigma_{\mnu}=0.08$ eV (same for + CMB or + CMB LE).}
\begin{center}
\begin{tabular}{ll cccc}
\hline\hline
Probes & prior & Planck & ACTpol & SPT & SPTpol \\ 
\hline

     $dN/dz+P(k)$      &     weak  & 1.97  & 2.43   & 1.18   & 1.06 \\ 
&         strong  & 2.34   & 3.36 & 1.67 & 1.66  \\ 
\hline
     $dN/dz+P(k)$ + CMB    &     weak  & 2.76 & 2.16  & 2.12 & 2.53 \\ 
&         strong  &3.25  & 2.53 & 2.33  & 2.88  \\ 
\hline
     $dN/dz+P(k)$ + CMB LE     &     weak  & 2.25  & 1.67  & 1.64  & 1.95 \\ 
&         strong  & 2.63  & 1.92  & 1.77  & 2.19  \\ 

 \hline
\end{tabular} 
\end{center}
\label{t:result_prior}
\end{table*}

\section{Discussion}
\label{sec:discussion}
\subsection{Parameter Degeneracies}
\label{sec:deg}
\begin{figure}
  \begin{center}
\includegraphics[width=0.45\textwidth]{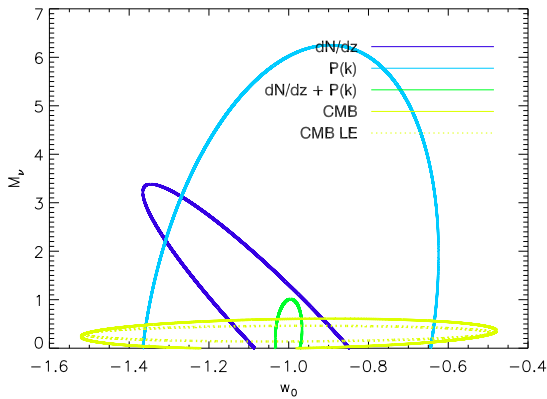} \\
       \caption{Joint constraints on the $\mnu$ and $w_0$. All curves denote $68\%$ confidence level, and are for number counts only (blue), power spectrum only (cyan), and combination of the two (green), Planck CMB (yellow), and Planck CMB LE (dotted yellow). }
     \label{fig:contour}
  \end{center}
\end{figure}

The dark energy equation of state $w_0$ and $\mnu$ is one of the major parameter degeneracies.~\reffig{contour} shows the $1\sigma$ constraints on $\mnu$ and $w_0$ computed from cluster number counts, power spectrum, combination of the two, with and without LE of the \planck\ CMB. The contour for number count shows a clear diagonal alignment, and the degeneracy direction can be understood as follows: an increase in neutrino mass suppresses the growth of structure formation, this can be compensated by a larger rate of accelerated expansion (i.e. more negative $w$). The constraints from power spectrum is less degenerate but show different degeneracy directions. As a result, combining information from both probes greatly improve the constraints. To see the effect of $w_0$ on $\mnu$ constraint, we derive $\sigma_{\mnu}$ again by marginalizing over  $w_0$ and $w_a$. We find that, as expected, only the constraints from number count are affected (improve by a factor of $>2$), while those from power spectrum are barely affected. Furthermore, only modest improvements are obtained when combining number count and power spectrum in this case. 

The degeneracy between curvature $\Omega_K$ and neutrino mass $\mnu$ is also known to be significant and impact on both $\mnu$ and the number of neutrino species $N_{eff}$, which could affect the constraints coming from CMB~\cite[e.g][]{Pierpaoli2003b,Smith2012}. 
However we note that the cluster probes used in this work are related to the growth of structures which are not sensitive to $\Omega_K$. Thus we expect that including $\Omega_K$ in the Fisher matrix analysis would not impact our results. It is out of the scope of this paper to study in depth the impacts of including an extended set of parameters (e.g. $\Omega_K$, $N_{eff}$). Nevertheless it would be potentially interesting to study their effects for growth of structures and we leave it for future works.


\subsection{Survey sensitivities to neutrino mass constraints}
In order to better understand what aspects of SZ surveys would improve the constraints on neutrino mass, we repeat the Fisher matrix calculation that includes different range of wavenumber $k$ and cluster mass $M$. 

\begin{figure*}
  \begin{center}
  \begin{tabular}{cc}
\includegraphics[width=0.50\textwidth]{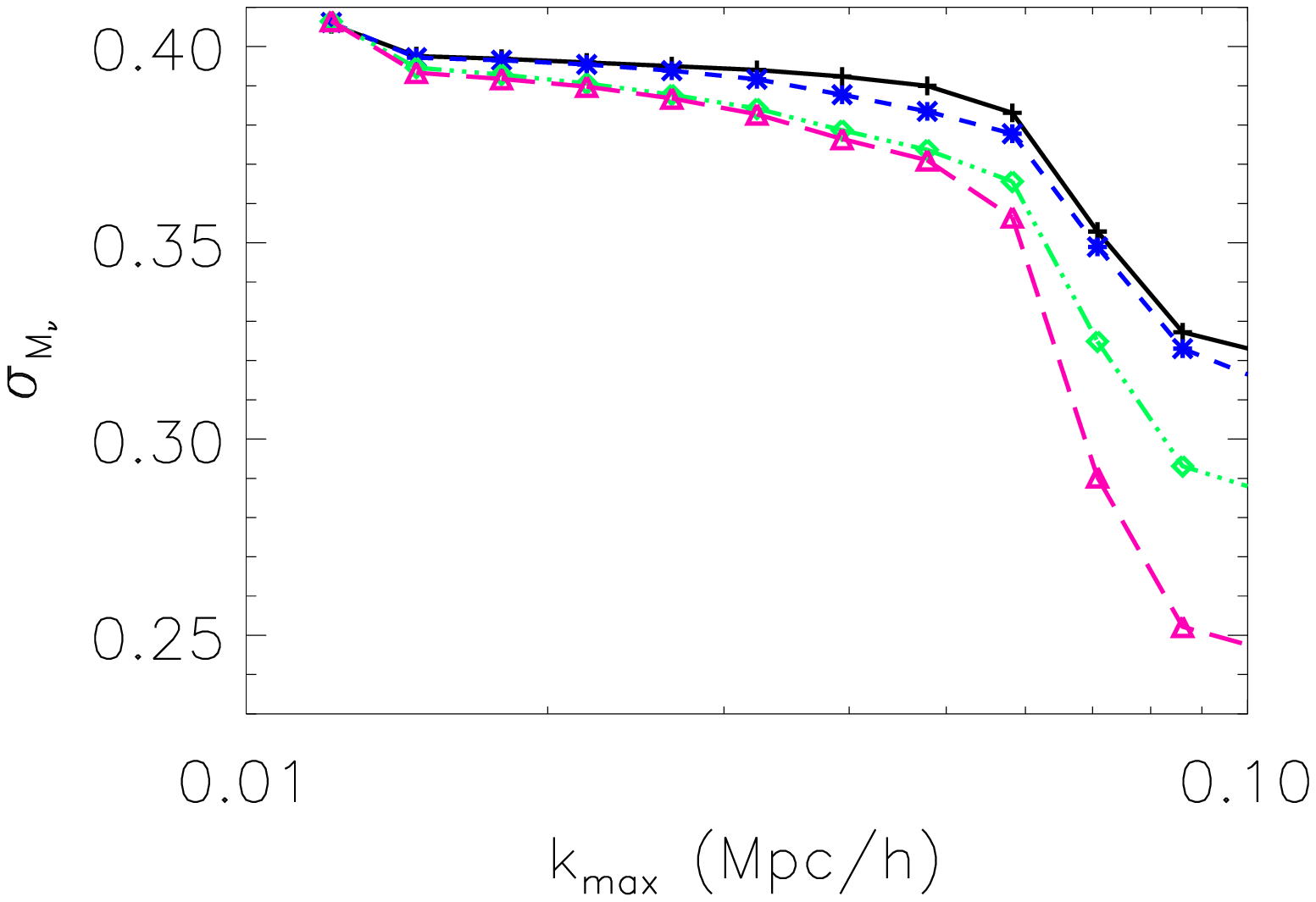}&
\includegraphics[width=0.48\textwidth]{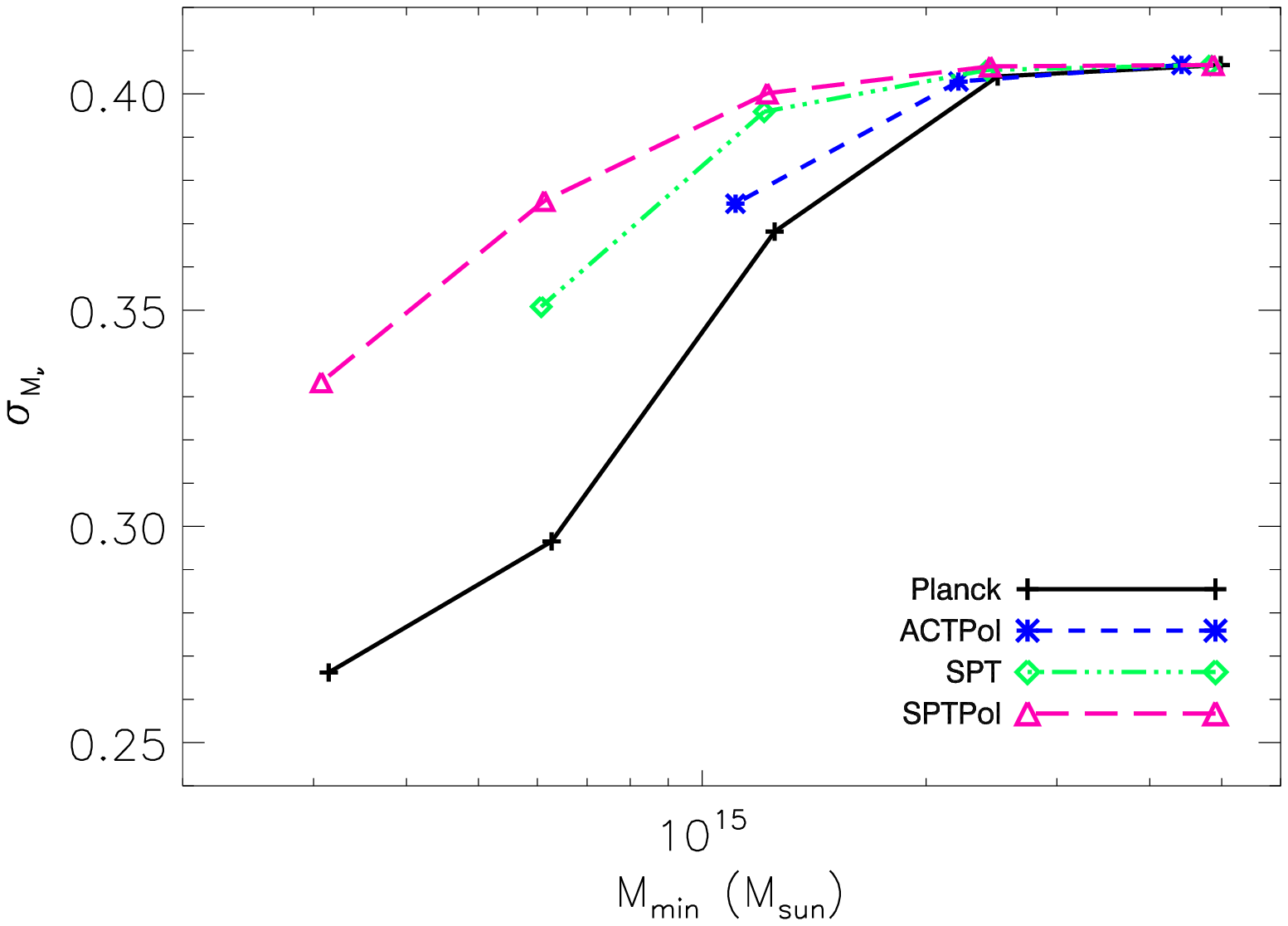} \\
\end{tabular}
       \caption{Fully marginalized constraints on $\mnu$ from the power spectrum of clusters + \planck\ CMB prior, as a function of maximum wavenumber $k_{\rm max}$ (left) and minimum cluster mass $M_{\rm min}$ (right). }
     \label{fig:sig}
  \end{center}
\end{figure*}

One of the major dependences is the maximum k values ($k_{\rm max}$) as it determines the smallest scales that can be probed by a survey. The effect of massive neutrino is particularly prominent at small scales (large $k$ values), in which free streaming of neutrinos prevent structure formation. In this work, we use the same $k_{\rm max}=0.1$ h/Mpc in the power spectrum Fisher matrix for all surveys considered. We do not attempt to increase the $k_{\rm max}$ beyond this value to avoid the non-linear effects at smaller scales. Furthermore, this scale is the limit that can be reached by SZ cluster surveys. Instead we study the dependence when small scale modes are lost, as shown in~\reffig{sig} (left). The effect of losing small scales information begins at $k\approx0.06$ h/Mpc, which corresponds to the free streaming scale $k_{\rm fs}$ at $z=2$, with $\mnu=0.3$ eV. The prospect of using smaller scale modes in constraining neutrino mass would be coming from galaxy surveys which can probe down to $k\approx0.5$ Mpc/h. A number of studies forecasted the neutrino mass constraints from future galaxy surveys (e.g.~\cite{Saito2009} (BOSS) and~\cite{Carbone2011} (EUCLID-like)) and found that the improvement in $\sigma_{\mnu}$ beyond $k=0.1$ Mpc/h is marginal (see Fig. 6 in~\cite{Saito2009}). 

The other relevant dependence is the limiting cluster mass that differentiates the various SZ surveys. We show in~\reffig{sig} (right) the fully marginalized $\sigma_{\mnu}$ from the power spectrum + \planck\ CMB as a function of minimum cluster mass used in the calculation. It is clear that a deeper survey that can probe down to lower mass would improve the constraints. It is also interesting to note that the SPTPol survey, despite of its small sky coverage, can perform better than the ACTPol and SPT survey because it can detect clusters down to $\approx2\times10^{14} M_{\odot}$. Therefore a deep survey can compensate for the sky coverage when constraining neutrino mass.


\section{Conclusion}
\label{sec:conclusion}
In this work, we explored the possibility of using future and upcoming SZ cluster surveys to constrain neutrino masses. We employ the Fisher Matrix analysis to forecast the sensitivities of various SZ surveys in constraining the total neutrino mass $\mnu$ in the context of flat $\Lambda$CDM cosmology. We do so by making use of the cluster number counts and power spectrum, and taking into account the self-calibration of mass-observable scaling relations.

In general, we find that the $\mnu$ constraints from cluster number count and power spectrum is weak if they are considered separately, due mainly to strong parameter degeneracy between $\mnu$ and $w_0$, especially in the case of cluster number count. However, such degeneracy can be broken if the two probes are combined, which helps to improve the constraints considerably. For example, a sample of $\approx1000$ clusters obtained from the \planck\ cluster survey gives $\sigma_{\mnu}=0.94$ eV (0.43 eV with weak prior). The constraints can be further improved when combined with CMB priors. The best constraint is obtained for the \planck\ and ACTPol survey, with $\sigma_{\mnu}=0.23$ eV (CMB) and $\sigma_{\mnu}=0.17$ eV (CMB LE). This is $\approx 80(25)\%$ improvement with respect to the CMB (CMB LE) only constraint. The use of CMB lensing extraction can better help the cluster only constraints because  it determines the neutrino's free streaming effect on the matter power spectrum and break some of the parameter degeneracies. 


While we find that $\mnu$ constraint is mainly driven by CMB and the addition of cluster probes, i.e. number count + power spectrum, to CMB only helps marginally, the use of clusters is still beneficial if we have good control of cluster systematics. For example, when applying a weak prior on the mass-observable relation, the $1\sigma$ error  on $\mnu$ as obtained from cluster count + power spectrum goes down to $0.48$ eV and is competitive with CMB only constraint. If we further combine with CMB priors, $\sigma_{\mnu}$ reduces to $0.07$ eV, which corresponds to a $\approx4\sigma$ detection for $\mnu\ge0.3$ eV.
The prospect of achieving the weak prior conditions is promising, e.g. clusters detected in weak lensing measurements or a subsample of objects having extensive multi-wavelength follow-up. Therefore, cluster measurements are useful, as an independent probe of the $\mnu$ with respect to the CMB, in tightening the current bound on $\mnu$.

We find that a deeper cluster survey that detects smaller mass clusters, e.g. down to $2\times10^{14} M_{\odot}$ like SPTPol, improves neutrino mass constraints. This is because of the effect of free streaming of massive neutrinos that prevents structure formation to happen at small scales. Likewise, the availability of the small scale modes, i.e. the maximum $k$ values that can be probed by a cluster survey, also helps the constraints. We show that the modes at $k\ge0.06$ h/Mpc are important as they help decreasing $\sigma_{\mnu}$ significantly.

\acknowledgments
EP acknowledges support from  JPL-Planck subcontract 1290790.
DM acknowledges support from USC Stauffer Fellowship. 
EP and DM were partially supported by  NASA grant NNX07AH59G.

\bibliography{ms}
\end{document}